\documentclass[authoryear,final,5p,times,twocolumn]{elsarticle}

 \usepackage{graphicx}

\usepackage{amssymb}

\journal{New Astronomy}

\def\aap{A\&A}
\def\aaps{A\&AS}

\def\apj{ApJ}

\def\apss{Ap\&SS}

\def\mnras{MNRAS}

\def\newa{NewA}
\def\nar{NewAR}
\def\araa{ARA\&A}
\def\astrobj#1{#1}

\begin{document}

\begin{frontmatter}



\title{Tentative insight into the multiplicity of the persistent dust maker WR\,106 from X-ray observations.\tnoteref{xmm}}
\tnotetext[xmm]{Based on observations with XMM-Newton, an ESA Science Mission with instruments and contributions directly funded by ESA Member states and the USA (NASA).}

\author{M. De Becker}
\ead{debecker@astro.ulg.ac.be}

\address{Department of Astrophysics, Geophysics and Oceanography, University of Li\`ege, 17, All\'ee du 6 Ao\^ut, B5c, B-4000 Sart Tilman, Belgium}

\begin{abstract}
This paper presents the results of the analysis of the very first dedicated X-ray observation with XMM-Newton of WR\,106. This carbon-rich WC9d Wolf-Rayet star belongs to the category of persistent dust makers (WCd stars). The issue of the multiplicity of these dust makers is pivotal to understand the dust formation process, and in this context X-ray observations may allow to reveal an X-ray emission attributable to colliding-winds in a binary system. The main result of this analysis is the lack of detection of X-rays coming from WR\,106. Upper limits on the X-ray flux are estimated, but the derived numbers are not sufficient to provide compelling constraints on the existence or not of a colliding-wind region. Detailed inspection of archive data bases reveals that persistent dust makers have been poorly investigated by the most sensitive X-ray observatories. Certainly, the combination of several approaches to indirectly constrain their multiplicity should be applied to lift a part of the veil on the nature of these persistent dust makers.
\end{abstract}


\end{frontmatter}

\section{Introduction}

Wolf-Rayet (WR) stars are known to be the evolved counterparts of O-type stars, with surface abundances typical of stellar nucleosynthesis in their stellar core \citep{crowther2007}. Among WR stars, the WC category includes evolved objects with enhanced carbon abundance and depleted nitrogen (CNO-cycle). WC stars produce very strong stellar winds, with mass loss rates up to two orders of magnitude larger than those of their O-type progenitors \citep{crowther2007}. Such dense -- and carbon-rich -- environments constitute privileged sites for the production of dust \citep{williams1995}. In particular, one may distinguish between persistent dust makers (with an apparently constant dust production rate) and episodic/periodic dust makers (with episodes of high dust production rate followed by more quiet activity regimes). The issue of the multiplicity of these objects has been raised very early, with for instance an intensive search for companions in a sample of WC stars by \citet{WV2000} using spectroscopic techniques. Even though in the case of episodic/periodic dust makers one is certainly dealing with binary systems with enhanced dust production activity close to periastron passage (in an eccentric orbit), the situation is not so clear for persistent dust makers. If the latter are also binaries, the absence of high amplitude modulation in the dust production may potentially point to almost circular systems. This category includes late-type WR stars, with sub-type WC9 along with a few WC8 objects \citep{williams1995,williams2014}. According to the classification introduced by \citet{vdh2001}, persistent dust makers are noted as WCd stars.

Among the category of persistent dust makers, the WC9d star WR\,106 (\astrobj{HD 313643}) is still poorly understood despite a significant effort devoted to this object in the past decades. It is one of the first WR stars which was identified as a dust maker thanks to the detection of a significant infrared excess \citep{ASH1972,CV1978}. Spectroscopic investigations of WR\,106 suggested a significant dilution of spectral lines pointing to the probable presence of a companion \citep{CK1977}, even though the study by \citet{torresconti1984} did not reveal the presence of absorption lines attributable to a potential nearby OB companion. The latter authors concluded that if there was a companion, its signature in the visible would be very weak. \citet{kato2002} reported on a deep transient fading in the optical light curve of WR\,106 but they refrained to interpret it in terms of a variability due to a companion. More recently, the in-depth photometric investigation by \citet{williams2014} revealed episodes of eclipse-like events reminiscent of the case of the pinwheel system WR\,104. It also worth to note that in the case of two other persistent dust makers, \astrobj{WR 69} and \astrobj{WR 104}, absorption lines probably attributable to a companion were discovered by \citet{WV2000}, lending support to the idea that the search for companions is relevant in this category of objects.

\begin{figure*}
\begin{center}
\includegraphics[width=150mm]{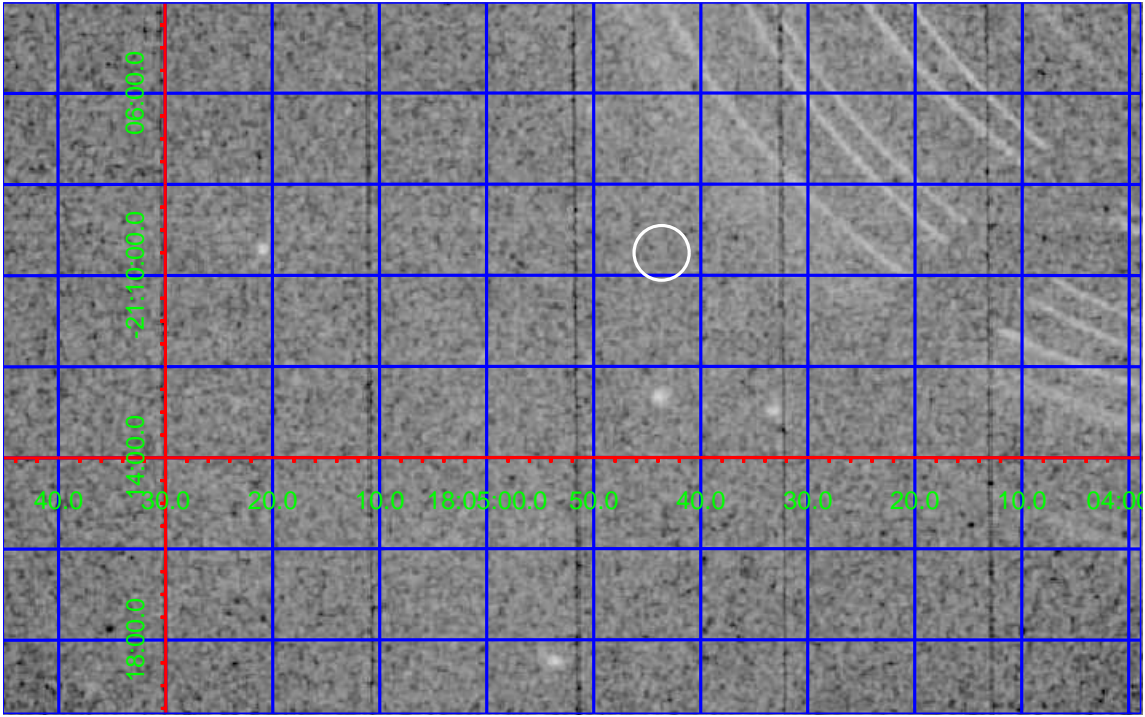}
\caption{EPIC-pn image obtained between 0.3 and 10.0\,keV. The circle indicates the position of WR106. A few point sources  with no relation with WR\,106 are present in the field of view. The North is up and the East is on the left.\label{pn}}
\end{center}
\end{figure*}

In this paper, we investigate the issue of the multiplicity of WR\,106 using X-ray observations. Single massive stars (OB and WR) are known to produce X-rays through thermal emission due to the presence of hot plasma heated by line driving instabilities in their stellar winds \citep{FeldX}. Single WC stars constitute an exception as they are not detected in X-rays, certainly because of a strong absorption of X-rays by the dense wind material \citep{oskinova2003}. In the case of binary systems, the colliding-winds offer an additional source of X-rays, generally with a harder emission spectrum than single stars \citep{SBP1992,PP2010}. X-rays can therefore provide hints for the presence of a companion. 

The investigation of the multiplicity of WR\,106 is therefore expected to benefit of the insight provided by X-rays. In Section\,\ref{obser}, we present the X-ray data sets, along with their processing. Section\,\ref{results} is devoted to a presentation of the results. Section\,\ref{disc} is dedicated to a discussion of WR\,106 in the wider context of other persistent dust makers. We finally conclude in Section\,\ref{concl}.

\section{Observations and data processing}\label{obser}

WR\,106 has been a target of the XMM-Newton satellite \citep{xmm} during the 11th Announcement of Opportunity (AO11, rev.\,2334), on 7th September, 2012 (Julian Date 2,455,896.375), under proposal ID\,069081 (PI: M. De Becker). EPIC instruments \citep{mos,pn} were operated in Full Frame mode, and the medium filter was used. These instruments are respectively the MOS and the pn cameras, allowing to obtain images and spectra within a field of view of about 30 arcmin diameter in the 0.3--10.0\,keV energy range. One given XMM-Newton observation allows to obtain simultaneously data with the pn and the two MOS cameras, each of them being at the focus of one of the three identical X-ray telescopes on-board the satellite. Beside some differences in the instrumental responses, data sets from these cameras differ mainly by the fact that photons collected by the telescope feeding the pn instrument are not shared with other instruments but the two other telescopes share their photons with MOS cameras and the RGS high-resolution spectrometers. This configuration explains why the sensitivity of the pn camera is better than that of MOS instruments. For details on these instruments, see notably the XMM-Newton Users Handbook\footnote{\tt http://xmm.esac.esa.int/external/xmm$\_$user$\_$support/documentation/uhb/index.html}. The exposure times were 22.4 and 20.7\,ks for MOS and pn instruments, respectively. Data were processed using the XMM-Newton Science Analysis Software (SAS) v.12.0.0 on the basis of the Observation Data Files (ODF) provided by the European Space Agency (ESA). Event lists were filtered using standard screening criteria (pattern $\leq$\,12 for MOS and pattern $\leq$\,4 for pn). The event lists were filtered to reject about 40\,\% of the exposure which was affected by a high background level due to a soft proton flare. 

The images in different energy bands show concentric arcs typical of straylight due to the presence of the bright Low-Mass X-ray Binary Sgr\,X-3 (RA = 18:01:32.3, DEC = $-$20:31:44, J2000) at about 1$^\circ$ from the position of WR\,106 (RA = 18:04:43.66, DEC = $-$21:09:30.5). The impact of straylight is mostly apparent to the North-West direction (see Figure\,1), and the position of WR\,106 does not seem to be substantially affected. The most striking result is the absence of any obvious point source at the position of WR\,106.

In order to investigate further the issue of the non detection of WR\,106 in X-rays, we used the data available at the High Energy Astrophysics Science Archive Research Center (HEASARC\footnote{http://heasarc.gsfc.nasa.gov}). Our target was in the field of 7 observations executed at the beginning of the 1990s with the ROentgen SATellite (ROSAT), as summarized in Table\,1. In this table, PSPCB and HRI stand respectively for the Position Sensitive Proportional Counter and High Resolution Imager instruments. In each case, the exposure time (Exp.) and the angular separation ($\delta$) between the position of WR\,106 and the centre of the field of view are specified. We retrieved all data sets and we inspected all images. In each case, we did not find any count excess likely to be attributable to a significant detection of WR\,106, consistently with the lack of detection with XMM which has a much better sensitivity. In the remaining of this paper, only the constraints on the X-ray emission estimated on the basis of XMM-Newton data will therefore be considered.

\begin{table}
\caption{Archive ROSAT observations including the position of WR\,106. \label{rosat}}
\begin{center}
\begin{tabular}{l c c c}
\hline
Seq.\,ID & Instr. & Exp. (s) & $\delta$ (') \\
\hline
\vspace*{-0.2cm}\\
rp500197n00 & PSPCB & 3248 & 20.35 \\
rp400282a01 & PSPCB & 3801 & 30.34 \\
rp400282n00 & PSPCB & 7432 & 30.34 \\
rp201060n00 & PSPCB & 9152 & 52.76 \\
rp900203n00 & PSPCB & 1974 & 57.92 \\
rh900391n00 & HRI & 6310 & 58.68 \\
rh900391a01 & HRI & 5642 & 58.68 \\
\vspace*{-0.2cm}\\
\hline
\end{tabular}
\end{center}
\end{table}

\section{Results from the XMM-Newton observation}\label{results}

\subsection{Upper limits}
We derived upper limits on the count rate between 0.3 and 10.0\,keV for the three EPIC instruments using the same approach as \citet{gossetwr40} and \citet{irc10420xmm}. We measured the number of counts (C) in a circular region (radius of 6'') centered on the expected position of the target, and we considered a count threshold (C$_\mathrm{max}$) corresponding to a logarithmic likelihood of 12. This translates into a probability to find a count number in excess of the critical value of about 6\,$\times$\,10$^{-6}$, under the null hypothesis of pure background fluctuations. The difference between these two quantities provides the count excess which was divided by the exposure time to yield the count rate evaluated within the extraction region (CR). Taking into account that the selected extraction region corresponds to an encircled energy fraction of the Point Spread Function of about 40\,\% (see the XMM User's Handbook), we finally derived the corrected count rates (CR$_\mathrm{cor}$) quoted in Table\,2 for all instruments.

\begin{figure*}[ht]
\begin{center}
\includegraphics[width=170mm]{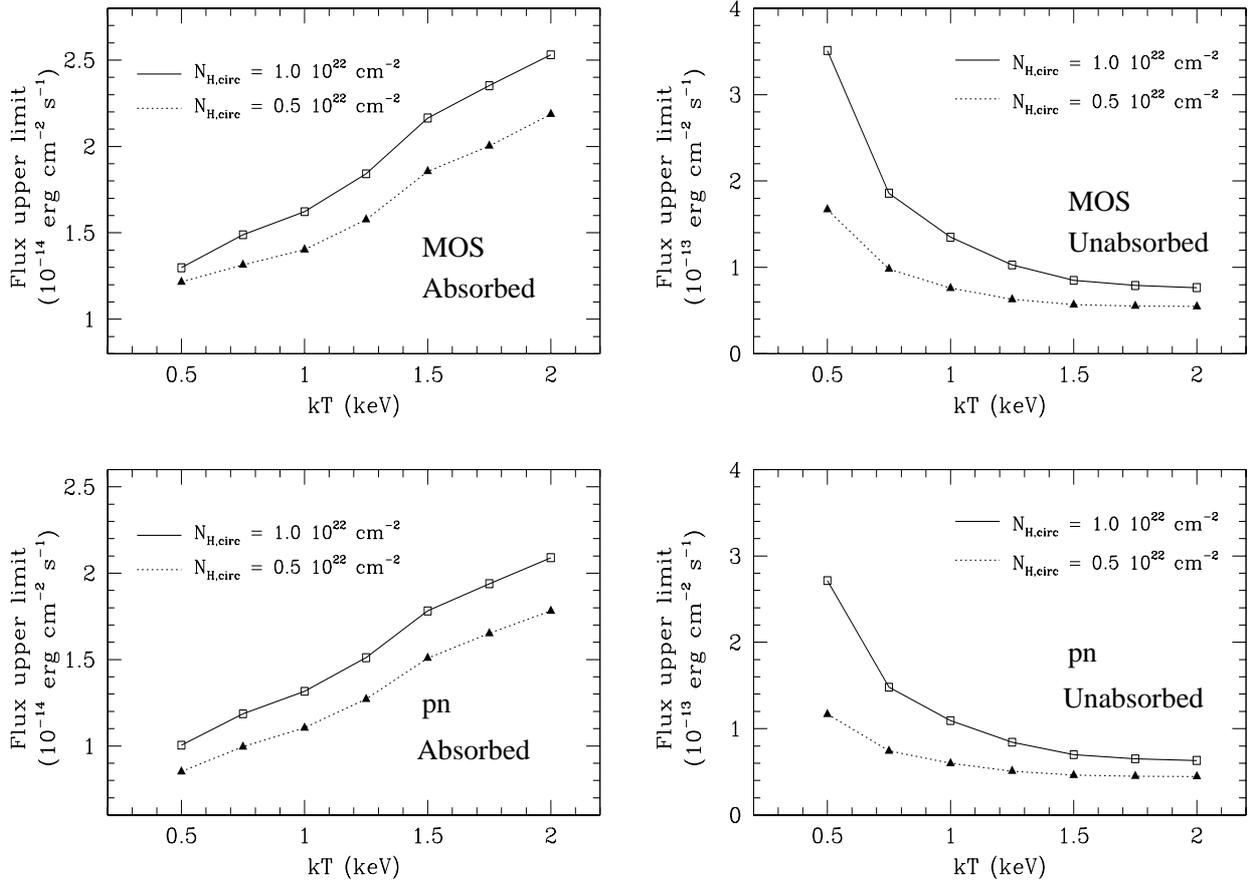}
\caption{Upper limits on the X-ray flux assuming a thermal emission model and two absorption columns (interstellar and circumstellar). {\it Left panels:} absorbed fluxes. {\it Right panels:} fluxes corrected for the interstellar absorption.\label{flux}}
\end{center}
\end{figure*}

The upper limits on the count rates were converted into fluxes (in erg\,cm$^{-2}$\,s$^{-1}$) using the WEBPIMMS\footnote{http://ledas-www.star.le.ac.uk/pimms/w3p/w3pimms.html} tool. An APEC optically thin thermal emission model affected by two absorption columns (interstellar and circumstellar) was assumed. For the interstellar absorption, we estimated the hydrogen column using the relation given by \citet{GL1989} and assuming $\mathrm{E(B-V)} = 1.2$ \citep{sander2012}, yielding N$_\mathrm{H,ISM}$ = 0.46\,$\times$\,10$^{22}$\,cm$^{-2}$. We defined a grid of models based on two values for the circumstellar absorbing column: N$_\mathrm{H,circ}$ = 0.5\,$\times$\,10$^{22}$\,cm$^{-2}$ and 1.0\,$\times$\,10$^{22}$\,cm$^{-2}$. These values are of the order of magnitude of wind column densities determined for some WR-type objects \citep[see e.g.][]{oskinova2003,wilwr140}. Extreme cases of wind absorption may motivate the selection of significantly larger values, but the result would be an complete extinction of X-rays from the object, whatever the nature and properties of the intrinsic spectrum. The latter situation would be relevant if the object is single, without any other source of X-rays away from the WR star such as a wind-wind interaction region. In a binary system, X-rays from the colliding-winds could escape in part the dense wind of the WC star, translating into a not so extreme value for the local absorbing column. The grid considers also seven plasma temperatures (kT between 0.5 and 2.0\,keV) spread over a range of values relevant for the thermal emission from early-type stars (single or binaries, see Sections\,\ref{singleX} and \ref{binaryX}). Our results are plotted in Figure\,2 for MOS and pn, either for absorbed fluxes or for fluxes corrected for the interstellar absorption.\\

\begin{table}
\caption{Estimates of the upper limits on the count rate for all EPIC instruments at the position of WR\,106.\label{counts}}
\begin{center}
\begin{tabular}{l c c c}
\hline
 & MOS1 & MOS2 & pn \\
\hline
\vspace*{-0.2cm}\\
C (cnt)& 5 & 5 & 24 \\
C$_\mathrm{max}$ $-$ C (cnt) & 17.9 & 17.9 & 48.7 \\
Eff. exp. time (s) & 14290 & 14860 & 11460 \\
CR (cnt\,s$^{-1}$) & 9.0\,$\times$\,10$^{-4}$ & 8.7\,$\times$\,10$^{-4}$ & 2.2\,$\times$\,10$^{-3}$ \\
CR$_\mathrm{cor}$ (cnt\,s$^{-1}$) & 2.3\,$\times$\,10$^{-3}$ & 2.2\,$\times$\,10$^{-3}$ & 5.4\,$\times$\,10$^{-3}$ \\
\vspace*{-0.2cm}\\
\hline
\end{tabular}
\end{center}
\end{table}

The trends shown by these plots deserve a few comments:
\begin{enumerate}
\item[1.] Upper limits derived from EPIC-MOS data are slightly higher than those derived on the basis of EPIC-pn, because of the better sensitivity of the latter instrument.
\item[2.] The absorbed fluxes increase with increasing plasma temperature of the emission model. Warmer plasma produce harder thermal spectra, with a significant fraction of their emission above the energies which are strongly affected by photoelectric absorption.
\item[3.] The model with the strongest circumstellar absorption leads to slightly higher upper limits. A given count number is made of higher energy photons when the source is more absorbed, as lower energy photons are the most affected. This results in a higher cumulated amount of energy measured in a given time interval. 
\item[4.] The ratio between fluxes corrected for the interstellar absorption and absorbed fluxes is much higher for soft emitting plasmas than for hard emission models. Once again, the stronger photoelectric absorption underwent by lower energy photons is substantially more significant when the bulk of the emission occurs at lower energies.
\end{enumerate}

The upper limits on the X-ray flux should be considered in the appropriate physical context, depending on the multiplicity status of WR\,106. Two scenarios will be envisaged: the {\it single star scenario} and the {\it binary scenario.}

\subsection{The single star scenario}\label{singleX}
The non detection of WR\,106 with XMM-Newton is reminiscent of the lack of detection of single WC-type stars in X-rays as discussed for instance by \citet{oskinova2003}. The absence of detected X-rays from WC stars is attributed to the very strong photoelectric absorption of X-ray photons by the dense stellar wind material, preventing X-rays produced in the inner layers of the wind to emerge from the outflowing envelope. This phenomenon is also responsible for the under-luminosity of evolved O-type supergiants \citep{debeckerofplusxmm,owocki2013}, and certainly for the non detection in X-rays of more evolved massive stars \citep{gossetwr40,irc10420xmm}.

In the context of the shocked wind scenario in single massive stars, the plasma temperature should be of the order of 0.5--0.7\,keV (i.e. 5--8\,$\times$\,10$^6$\,K). The post-shock temperatures are associated to pre-shock velocities of a few hundreds km\,s$^{-1}$, in agreement with the typical relative velocities of out-moving shells of material colliding in individual stellar winds (see e.g. \citealt{FeldX}). In Figure\,2, the single star scenario is therefore more specifically reproduced by the left part of the plots. This translates into absorbed flux upper limits not higher than 1.5\,$\times$\,10$^{-14}$\,erg\,cm$^{-2}$\,s$^{-1}$. Upper limits corrected for the interstellar absorption spread over a range between 1 and 4\,$\times$\,10$^{-13}$\,erg\,cm$^{-2}$\,s$^{-1}$, translating into upper limits on the X-ray luminosity (L$_\mathrm{X}$) of 1.1--4.5\,$\times$\,10$^{32}$\,erg\,s$^{-1}$ (assuming a distance of 3.06\,kpc, \citealt{sander2012}). Considering the bolometric luminosity (L$_\mathrm{bol}$\,=\,1.43\,$\times$\,10$^5$\,L$_\odot$\,=\,5.58\,$\times$\,10$^{38}$\,erg\,s$^{-1}$) given by \citet{sander2012}, the upper limit on the L$_\mathrm{X}$/L$_\mathrm{bol}$ ratio is 2--8\,$\times$\,10$^{-7}$ (depending on the assumption on the plasma temperature and the circumstellar absorbing column). These upper limits are of the order of magnitude of -- or larger than -- the expected emission from single regular O-type stars \citep{sanalxlbol,owocki2013}. Our upper limits suggest that, as a single star, WR\,106 emits less -- or a similar amount of -- X-rays than O-type stars. This is not unexpected as we are dealing with an evolved object with a much denser stellar wind which absorbs efficiently intrinsically-produced X-rays. The non detection of WR\,106 agrees therefore with the single WC-type star scenario.

\subsection{The binary scenario}\label{binaryX}
In the context of a wind-wind interaction in a binary system made of a WC star with an OB companion, pre-shock velocities are expected to be significantly larger than in the case of intrinsic shocks of individual stellar winds (see Section\,\ref{singleX}). In short period systems (typically a few days) the stellar winds will interact in their acceleration zone (typically a few stellar radii). For longer period systems, stellar winds will likely reach their terminal velocities before colliding, leading therefore to hotter plasma with temperatures higher than 10$^7$\,K ($>$\,1\,keV)\footnote{Note that the particular case of very unequal winds with the strongest one crashing onto the surface of the companion -- with complete disruption of the wind-wind interaction region -- is not considered here. Such a scenario is for instance discussed by \citet{ParkinSim2013}.}. In the grid of models considered in Figure\,2, we will therefore focus on the right part of the plots. For our range of assumed parameters we derive upper limits on the absorbed flux of about 1.0--2.5\,$\times$\,10$^{-14}$\,erg\,cm$^{-2}$\,s$^{-1}$, and of about 0.5--1.5\,$\times$\,10$^{-13}$\,erg\,cm$^{-2}$\,s$^{-1}$ for fluxes corrected for the interstellar absorption. The latter values convert into L$_\mathrm{X}$ of 0.6--1.7\,$\times$\,10$^{32}$\,erg\,s$^{-1}$ and a L$_\mathrm{X}$/L$_\mathrm{bol}$ ratio of 1--3\,$\times$\,10$^{-7}$. If WR\,106 was producing X-rays in a wind-wind interaction region, it would be expected to be characterized by a L$_\mathrm{X}$/L$_\mathrm{bol}$ ratio significantly larger than 10$^{-7}$. This strongly suggests that no colliding-winds are strongly contributing to the X-ray emission of WR\,106, otherwise it would likely have been detected with XMM-Newton.

However, one should keep in mind that the infrared excess reported for WR\,106 -- in relation with its dust production -- leads to the existence of a circumstellar envelope likely to absorb a significant amount of X-rays potentially produced in a putative colliding-wind system. Consequently, the derived upper limits do not allow to reject definitely the binary scenario.

\section{Discussion}\label{disc}

\subsection{Other persistent dust makers in X-rays}\label{other}
It is known that some variable dust makers have already been intensively observed in X-rays. The most striking example is certainly \astrobj{WR 140} \citep[see e.g.][]{wilwr140,pollock2005}. However, the case of WR\,106 deserves to be discussed in the context of other stars belonging to the same category, namely that of persistent (stable) dust makers. Paul Crowther's Galactic Wolf-Rayet Catalogue\footnote{http://pacrowther.staff.shef.ac.uk/WRcat/} includes 37 objects with the WC9d classification. Among the most studied WCd stars so far, one could mention the sample of 16 objects investigated by \citet{williams2014} in the infrared. The XMM-Newton archives were carefully inspected to search for data sets including some of these objects. Relevant EPIC exposures were found in the cases of \astrobj{WR 65} and \astrobj{WR 96}. WR\,65 is known as a binary system, with an unidentified OB companion. Nine exposures including this object in the field of view were found in the XMM-Newton archive database, with exposure times of 10--30\,ks. The Wolf-Rayet system does not seem to be detected, in spite of its confirmed binary nature expected to lead to some additional X-rays produced by the colliding-winds. The X-ray images are dominated by the very X-ray bright supernova remnant G320.4-1.2\,N, preventing any detailed inspection of that part of the field. The position of WR\,96 is located close to the outer boundary of the EPIC field of four exposures of about 20--30\,ks and it is not detected.

The same exercise using the Chandra archive database led to the following findings. WR\,65 (see above) was also found in the field of Chandra exposures, with no detection. This is not a surprise considering the better sensitivity of XMM-Newton. In addition, a very short exposure of about 3\,ks was found for \astrobj{WR 121}, certainly much too short to expect any detection of such an object anticipated to be faint in X-rays.

It seems therefore that the current census of observations of WCd stars with the most recent X-ray facilities is quite poor, lending further relevance to the present dedicated observation of WR\,106. Probably the category of WCd stars includes objects much too faint to be reached by present X-ray observatories. However, additional dedicated exposures should be organized to clarify this idea. Considering the lack of such observations so far, a few additional ones could constitute a significant improvement with respect to the present situation.

\subsection{The multiplicity of WCd stars}\label{mult}
The infrared photometric investigation by \citet{williams2014} revealed a large number of eclipse-like events for WR\,106, presenting some similarities with WR\,104. The latter system is known as a binary system with a period of about 240 days \citep{tuthill2008}. Such eclipse-like events could be attributed to clumps of material crossing the line of sight with a recurrence time depending on their distribution. The other WCd stars investigated by \citet{williams2014} present only a few eclipse-like events over several decades, suggesting significant differences between the wind properties of WR\,104 and WR\,106 on the one hand, and the other WCd stars on the other hand. These different behaviours point to a potential strong difference in the clump formation, and in particular of their distribution. A possible way to reconcile these discrepancies proposed by \citet{williams2014} is the existence of some anisotropy in the stream of clumps, with significant differences coming from different viewing angles. 

The existence of the infrared excess and the occurrence of the eclipses point clearly to the presence of circumstellar material likely to significantly absorb X-rays coming from the central object (either single or binary). If one wants to privilege a unique picture, it is relevant to emphasize a few facts:
\begin{enumerate}
\item[-] all WCd stars investigated on the basis of long-term infrared photometric time series display eclipses most probably due to clumps in the circumstellar regions,
\item[-] no WCd has been detected in X-rays so far,
\item[-] a few WCd stars have been identified as binary systems. 
\end{enumerate}

Let us assume, first, that all WCd stars are binaries. In this case, the dust production should proceed thanks to the enhanced density regions resulting from the colliding-winds, in locations efficiently shielded from the WC photospheric light \citep{williams1995,cherchneff2000,crowther2003}. The colliding-wind region would be expected to produce some thermal X-rays, but the circumstellar material would be opaque enough to prevent any detection with the most recent X-ray observatories (at least in the few cases investigated so far). This extinction of X-rays produced by the colliding-wind region could be more easily explained if the stellar separation is not too large. A wide system would indeed allow hopefully some X-rays to escape the dusty envelope. On the other hand, the escape of X-rays would be strongly dependent on the orientation of the system, opening the possibility that at some orbital phases a weak X-ray signal could be detected. In this binary system scenario, several hidden companions would wait to be detected. However, the difficulty to detect them would be twofold: (i) spectroscopic investigations would require to detect and monitor spectral features from the secondary, certainly significantly fainter than the WC star, and (ii) high angular resolution astrometric investigations are generally conducted in the infrared domain, and would therefore require to detect a faint companion in a nebula populated by dust particles bright in the infrared. Hints for the existence of a companion may also be provided by the detection of synchrotron radio emission, explained by the existence of relativistic electrons accelerated by the shocks in the colliding-wind region in a binary system. Among persistent dust makers, \astrobj{WR 98a}, WR\,104 and WR\,112\footnote{Direct evidence for the existence of a companion in the case of WR\,112 is still lacking.} are known to belong to this category of particle-accelerating colliding-wind binaries \citep{pacwbcata}. Beside that, it is worth to mention that other WCd objects are known to be binaries, such as WR\,65 \citep{vdh2001} and \astrobj{W239} \citep{clark2011}.

Let us assume now that some WCd stars might be single objects, in addition to the few ones known to be binaries. It would justify immediately the lack of X-ray detection because of the strong local absorption by the stellar wind material, which is a general property of single WC stars \citep{oskinova2003}. In this context, the circumstellar dust would just constitute a complementary source of X-ray extinction, in addition to that attributed to the dense WC wind material in general. The clumps responsible for the eclipse revealed by photometric investigations would be produced by the stellar winds of the WC star, without the need of a participation of any colliding-wind region. This would therefore require that dust production and formation of clumps on the one hand, and binarity along with the formation of a few pinwheel nebulae on the other hand, may not necessarily be intimately related. In this scenario, dust production by colliding-winds may take place on top of the intrinsic production in the WC wind.

\section{Concluding remarks}\label{concl}
A dedicated XMM-Newton observation of the persistent dust maker WR\,106 was analyzed in details. The main result stands
in the lack of detection of the WC9d star. Upper limits on the count rates were derived. These limits were converted into upper limits on the X-ray flux (in physical units) on the basis of adequate models relevant for the X-ray emission from massive stars. These limits were discussed in the context of both scenarios: single star or binary system. However, the collected information could not discriminate between these two likely scenarios.

It appears also that the inspection of persistent dust makers (belonging to the WCd class) in X-rays is quite poor, with only a few systems appearing in the field of view of modern observatories. No WCd object seems to be detected in X-rays so far, including the known binary WR\,65. A more extended inspection of this class of objects is expected to clarify our view of persistent dust makers, notably to constrain further the X-ray emission that is expected to arise from the colliding-wind region, provided they are binaries. In addition, in the binary scenario, some variations in the absorption of X-rays could be expected as a function of the orbital phase, allowing to expect a changing detection probability depending on the geometry of the system.

The question is still: do we need a binary system to persistently produce dust, or can it be produced in single WC-star winds? On the basis of the current census of observational information, both scenarios are still in competition. Considering the high anticipated difficulties to detect putative companions in such systems, certainly indirect approaches such as dedicated X-ray observations, high dynamic range and high resolution interferometry and radio observations aiming at identifying synchrotron radiation should be combined to improve our view of their multiplicity. 

\section*{Acknowledgments}

The author wants to warmly thank people working at the XMM-SOC for the scheduling of the XMM-Newton observations, along with the anonymous referee for constructive comments on the manuscript. The SIMBAD database has been consulted for the bibliography.

\bibliographystyle{elsarticle-harv}

\begin{thebibliography}{29}
\expandafter\ifx\csname natexlab\endcsname\relax\def\natexlab#1{#1}\fi
\expandafter\ifx\csname url\endcsname\relax
  \def\url#1{\texttt{#1}}\fi
\expandafter\ifx\csname urlprefix\endcsname\relax\def\urlprefix{URL }\fi

\bibitem[{{Allen} et~al.(1972){Allen}, {Swings}, and {Harvey}}]{ASH1972}
{Allen}, D.~A., {Swings}, J.~P., {Harvey}, P.~M., 1972. {Infrared photometry of
  northern Wolf-Rayet stars.} \aap 20, 333--336.

\bibitem[{{Cherchneff} et~al.(2000){Cherchneff}, {Le Teuff}, {Williams}, and
  {Tielens}}]{cherchneff2000}
{Cherchneff}, I., {Le Teuff}, Y.~H., {Williams}, P.~M., {Tielens}, A.~G.~G.~M.,
  May 2000. {Dust formation in carbon-rich Wolf-Rayet stars. I. Chemistry of
  small carbon clusters and silicon species}. \aap 357, 572--580.

\bibitem[{{Clark} et~al.(2011){Clark}, {Ritchie}, {Negueruela}, {Crowther},
  {Damineli}, {Jablonski}, and {Langer}}]{clark2011}
{Clark}, J.~S., {Ritchie}, B.~W., {Negueruela}, I., {Crowther}, P.~A.,
  {Damineli}, A., {Jablonski}, F.~J., {Langer}, N., Jul. 2011. {A VLT/FLAMES
  survey for massive binaries in Westerlund 1. III. The WC9d binary W239 and
  implications for massive stellar evolution}. \aap 531, A28.

\bibitem[{{Cohen} and {Kuhi}(1977)}]{CK1977}
{Cohen}, M., {Kuhi}, L.~V., Jul. 1977. {Wolf-Rayet stars. VII - Optical
  spectropolarimetry of WC9 stars}. \mnras 180, 37--44.

\bibitem[{{Cohen} and {Vogel}(1978)}]{CV1978}
{Cohen}, M., {Vogel}, S.~N., Oct. 1978. {Wolf-Rayet stars - VIII. 2- to
  4-$\mu$m spectrophotometry of late WC stars.} \mnras 185, 47--55.

\bibitem[{{Crowther}(2003)}]{crowther2003}
{Crowther}, P.~A., 2003. {Dust Formation around Wolf-Rayet Stars}. \apss 285,
  677--685.

\bibitem[{{Crowther}(2007)}]{crowther2007}
{Crowther}, P.~A., Sep. 2007. {Physical Properties of Wolf-Rayet Stars}. \araa
  45, 177--219.

\bibitem[{{De Becker}(2013)}]{debeckerofplusxmm}
{De Becker}, M., Dec. 2013. {The X-ray under-luminosity of the O-type
  supergiants HD 16691 and HD 14947 revealed by XMM-Newton}. \newa 25, 7--11.

\bibitem[{{De Becker} et~al.(2014){De Becker}, {Hutsem{\'e}kers}, and
  {Gosset}}]{irc10420xmm}
{De Becker}, M., {Hutsem{\'e}kers}, D., {Gosset}, E., May 2014. {The XMM-Newton
  view of the yellow hypergiant IRC + 10420 and its surroundings}. \newa 29,
  75--81.

\bibitem[{{De Becker} and {Raucq}(2013)}]{pacwbcata}
{De Becker}, M., {Raucq}, F., Oct. 2013. {Catalogue of particle-accelerating
  colliding-wind binaries}. \aap 558, A28.

\bibitem[{{Feldmeier} et~al.(1997){Feldmeier}, {Puls}, and {Pauldrach}}]{FeldX}
{Feldmeier}, A., {Puls}, J., {Pauldrach}, A.~W.~A., Jun. 1997. {A possible
  origin for X-rays from O stars.} \aap 322, 878--895.

\bibitem[{{Gosset} et~al.(2005){Gosset}, {Naz{\'e}}, {Claeskens}, {Rauw},
  {Vreux}, and {Sana}}]{gossetwr40}
{Gosset}, E., {Naz{\'e}}, Y., {Claeskens}, J.-F., {Rauw}, G., {Vreux}, J.-M.,
  {Sana}, H., Jan. 2005. {An XMM-Newton look at the Wolf-Rayet star WR 40. The
  star itself, its nebula and its neighbours}. \aap 429, 685--704.

\bibitem[{{Groenewegen} and {Lamers}(1989)}]{GL1989}
{Groenewegen}, M.~A.~T., {Lamers}, H.~J.~G.~L.~M., Sep. 1989. {The winds of
  O-stars. I - an analysis of the UV line profiles with the SEI method}. \aaps
  79, 359--383.

\bibitem[{{Jansen} et~al.(2001){Jansen}, {Lumb}, {Altieri}, {Clavel}, {Ehle},
  {Erd}, {Gabriel}, {Guainazzi}, {Gondoin}, {Much}, {Munoz}, {Santos},
  {Schartel}, {Texier}, and {Vacanti}}]{xmm}
{Jansen}, F., {Lumb}, D., {Altieri}, B., {Clavel}, J., {Ehle}, M., {Erd}, C.,
  {Gabriel}, C., {Guainazzi}, M., {Gondoin}, P., {Much}, R., {Munoz}, R.,
  {Santos}, M., {Schartel}, N., {Texier}, D., {Vacanti}, G., Jan. 2001.
  {XMM-Newton observatory. I. The spacecraft and operations}. \aap 365, L1--L6.

\bibitem[{{Kato} et~al.(2002){Kato}, {Haseda}, {Takamizawa}, and
  {Yamaoka}}]{kato2002}
{Kato}, T., {Haseda}, K., {Takamizawa}, K., {Yamaoka}, H., Oct. 2002. {Deep
  transient optical fading in the WC9 Star WR 106}. \aap 393, L69--L71.

\bibitem[{{Oskinova} et~al.(2003){Oskinova}, {Ignace}, {Hamann}, {Pollock}, and
  {Brown}}]{oskinova2003}
{Oskinova}, L.~M., {Ignace}, R., {Hamann}, W.-R., {Pollock}, A.~M.~T., {Brown},
  J.~C., May 2003. {The conspicuous absence of X-ray emission from
  carbon-enriched Wolf-Rayet stars}. \aap 402, 755--765.

\bibitem[{{Owocki} et~al.(2013){Owocki}, {Sundqvist}, {Cohen}, and
  {Gayley}}]{owocki2013}
{Owocki}, S.~P., {Sundqvist}, J.~O., {Cohen}, D.~H., {Gayley}, K.~G., Mar.
  2013. {Thin-shell mixing in radiative wind-shocks and the L$_{x}$ $\sim$
  L$_{bol}$ scaling of O-star X-rays}. \mnras 429, 3379--3389.
  
\bibitem[{{Parkin} and {Sim}(2013)}]{ParkinSim2013}
{Parkin}, E.~R., {Sim}, S.~A., Apr. 2013. {Self-regulated Shocks in Massive Star Binary Systems}. \apj 767, 114.  

\bibitem[{{Pittard} and {Parkin}(2010)}]{PP2010}
{Pittard}, J.~M., {Parkin}, E.~R., Apr. 2010. {3D models of radiatively driven
  colliding winds in massive O + O star binaries - III. Thermal X-ray
  emission}. \mnras 403, 1657--1683.
  
\bibitem[{{Pollock} et al.(2005){Pollock},{Corcoran},{Stevens}, and {Williams}}]{pollock2005}
 {Pollock}, A.~M.~T., {Corcoran}, M.~F., {Stevens}, I.~R. and {Williams}, P.~M., Aug. 2005. {Bulk Velocities, Chemical Composition, and Ionization Structure of the X-Ray Shocks in WR 140 near Periastron as Revealed by the Chandra Gratings}. \apj 629, 482--498.

\bibitem[{{Sana} et~al.(2006){Sana}, {Rauw}, {Naz{\'e}}, {Gosset}, and
  {Vreux}}]{sanalxlbol}
{Sana}, H., {Rauw}, G., {Naz{\'e}}, Y., {Gosset}, E., {Vreux}, J.-M., Oct.
  2006. {An XMM-Newton view of the young open cluster NGC 6231 - II. The OB
  star population}. \mnras 372, 661--678.

\bibitem[{{Sander} et~al.(2012){Sander}, {Hamann}, and {Todt}}]{sander2012}
{Sander}, A., {Hamann}, W.-R., {Todt}, H., Apr. 2012. {The Galactic WC stars.
  Stellar parameters from spectral analyses indicate a new evolutionary
  sequence}. \aap 540, A144.

\bibitem[{{Stevens} et~al.(1992){Stevens}, {Blondin}, and {Pollock}}]{SBP1992}
{Stevens}, I.~R., {Blondin}, J.~M., {Pollock}, A.~M.~T., Feb. 1992. {Colliding
  winds from early-type stars in binary systems}. \apj 386, 265--287.

\bibitem[{{Str{\"u}der} et~al.(2001){Str{\"u}der}, {Briel}, {Dennerl},
  {Hartmann}, {Kendziorra}, {Meidinger}, and {Pfeffermann}}]{pn}
{Str{\"u}der}, L., {Briel}, U., {Dennerl}, K., {Hartmann}, R., {Kendziorra},
  E., {Meidinger}, N., {Pfeffermann}, E. e.~a., Jan. 2001. {The European Photon
  Imaging Camera on XMM-Newton: The pn-CCD camera}. \aap 365, L18--L26.

\bibitem[{{Torres} and {Conti}(1984)}]{torresconti1984}
{Torres}, A.~V., {Conti}, P.~S., May 1984. {The spectra of Wolf-Rayet stars. II
  - The WC 9 subclass}. \apj 280, 181--188.

\bibitem[{{Turner} et~al.(2001){Turner}, {Abbey}, {Arnaud}, {Balasini},
  {Barbera}, {Belsole}, and {Bennie}}]{mos}
{Turner}, M.~J.~L., {Abbey}, A., {Arnaud}, M., {Balasini}, M., {Barbera}, M.,
  {Belsole}, E., {Bennie}, P.~J. e.~a., Jan. 2001. {The European Photon Imaging
  Camera on XMM-Newton: The MOS cameras}. \aap 365, L27--L35.

\bibitem[{{Tuthill} et~al.(2008){Tuthill}, {Monnier}, {Lawrance}, {Danchi},
  {Owocki}, and {Gayley}}]{tuthill2008}
{Tuthill}, P.~G., {Monnier}, J.~D., {Lawrance}, N., {Danchi}, W.~C., {Owocki},
  S.~P., {Gayley}, K.~G., Mar. 2008. {The Prototype Colliding-Wind Pinwheel WR
  104}. \apj 675, 698--710.

\bibitem[{{van der Hucht}(2001)}]{vdh2001}
{van der Hucht}, K.~A., Feb. 2001. {The VIIth catalogue of galactic Wolf-Rayet
  stars}. \nar 45, 135--232.

\bibitem[{{Williams}(1995)}]{williams1995}
{Williams}, P.~M., 1995. {Dust formation around WC stars (Invited)}. In: {van
  der Hucht}, K.~A., {Williams}, P.~M. (Eds.), Wolf-Rayet Stars: Binaries;
  Colliding Winds; Evolution. Vol. 163 of IAU Symposium. p. 335.

\bibitem[{{Williams}(2014)}]{williams2014}
{Williams}, P.~M., Dec. 2014. {Eclipses and dust formation by WC9 type
  Wolf-Rayet stars}. \mnras 445, 1253--1260.

\bibitem[{{Williams} and {van der Hucht}(2000)}]{WV2000}
{Williams}, P.~M., {van der Hucht}, K.~A., May 2000. {Spectroscopy of WC9
  Wolf-Rayet stars: a search for companions}. \mnras 314, 23--32.
  
 \bibitem[{{Williams} et~al.(1990){Williams},{van der Hucht},{Pollock},{Florkowski},{van der Woerd}, and {Wamsteker}}]{wilwr140} 
{Williams}, P.~M., {van der Hucht}, K.~A., {Pollock}, A.~M.~T., {Florkowski}, D.~R., {van der Woerd}, H., {Wamsteker}, W.~M., Apr. 1990. {Multi-frequency variations of the Wolf-Rayet system HD 193793. I - Infrared, X-ray and radio observations}. \mnras 243, 662--684.

\end{thebibliography}

\end{document}